# Deep Learning Algorithms for Early Diagnosis of Acute Lymphoblastic Leukemia

Dimitris Papaioannou, Ioannis T. Christou

The American College of Greece

Nikos Anagnou

Dept. of Medicine, National & Kapodistrean University of Athens and Biomedical Research Foundation, Academy of Athens, Greece

Aristotelis Hadjiioannou

Biomedical Research Foundation, Academy of Athens, Greece

## Abstract

Acute lymphoblastic leukemia (ALL) is a form of blood cancer that affects the white blood cells. ALL constitutes approximately 25% of pediatric cancers. Early diagnosis and treatment of ALL are crucial for improving patient outcomes. The task of identifying immature leukemic blasts from normal cells under the microscope can prove challenging, since the images of a healthy and cancerous cell appear similar morphologically. In this study, we propose a binary image classification model to assist in the diagnostic process of ALL. Our model takes as input microscopic images of blood samples and outputs a binary prediction of whether the sample is normal or cancerous. Our dataset consists of 10661 images out of 118 subjects. Deep learning techniques on convolutional neural network architectures were used to achieve accurate classification results. Our proposed method achieved 94.3% accuracy and could be used as an assisting tool for hematologists trying to predict the likelihood of a patient developing ALL.

## Introduction

### Introduction to the Clinical Problem at Hand

Acute lymphoblastic leukemia (ALL) is a type of cancer that affects the white blood cells, known as lymphocytes. It is the most common type of childhood cancer, and it can also occur in adults. ALL is characterized by the rapid and uncontrolled growth of immature white blood cells, known as lymphoblasts, which can crowd out healthy cells and prevent the body from fighting the infection. The clinical problem with ALL is that it can progress quickly and aggressively, and if left untreated, can prove fatal. The symptoms of ALL can be similar to those of other, less serious conditions, which can make it difficult to diagnose. Symptoms can include fatigue, weakness, fever, weight loss, and easy bruising or bleeding.

Diagnosis of ALL typically involves a combination of blood tests, bone marrow biopsy, and imaging studies. However, diagnosing ALL can be challenging for several reasons:

- Symptoms are often non-specific: The symptoms of ALL can be similar to those of other, less serious conditions, such as the flu or anemia, which can make it difficult to diagnose.
- Lack of specific diagnostic markers: There are currently, no specific diagnostic markers for ALL, which makes it difficult to distinguish from other types of leukemia or other hematological disorders.
- Variation in the presentation: ALL can appear in different ways, and it can vary depending on the age of the patient and the subtype of ALL. This can make the diagnosis difficult, especially in cases where the disease is less advanced.
- Limited access to specialized testing: In some areas, access to specialized testing and facilities, such as bone marrow biopsy, may be limited, thus rendering the diagnosis of ALL more challenging.
- Differential diagnosis: ALL often needs to be differentiated from other types of leukemia as well as other hematological disorders. This can be challenging and requires a thorough evaluation.

Overall, the diagnosis of ALL requires a combination of clinical, laboratory and imaging studies, along with the expertise of a hematologist. Even with all the available tools, the diagnosis of ALL can still be challenging.

# Review of Machine Learning Methods for the Problem at Hand

An arsenal of Machine learning techniques can be utilized to analyze medical data, derive patterns, and make predictions, utterly assisting hematologists in several ways.
Machine learning algorithms can be trained to analyze images of blood cells and identify abnormal cells, which can aid in the diagnosis of blood disorders such as leukemia. Trained on a large scale of patient data, such as medical history and lab results, these algorithms can also assist in the prediction of which patients are at risk of certain blood cancers, allowing for early diagnosis and intervention. The latter can prove crucial for effective treatment.

## Convolutional Neural Networks

Convolutional Neural Networks (CNNs) are a powerful tool for image classification tasks and have been widely used in the field of cell image classification. They are able to learn complex features from the images and can achieve high accuracy rates. One of the main advantages of CNNs is that they can automatically learn the features from the images which reduces the need of manual feature extraction. This is particularly useful when dealing with microscope cell images, which can be complex and vary widely in appearance. CNNs are also able to capture the spatial relationships between the cells, which is important for accurate cell classification.

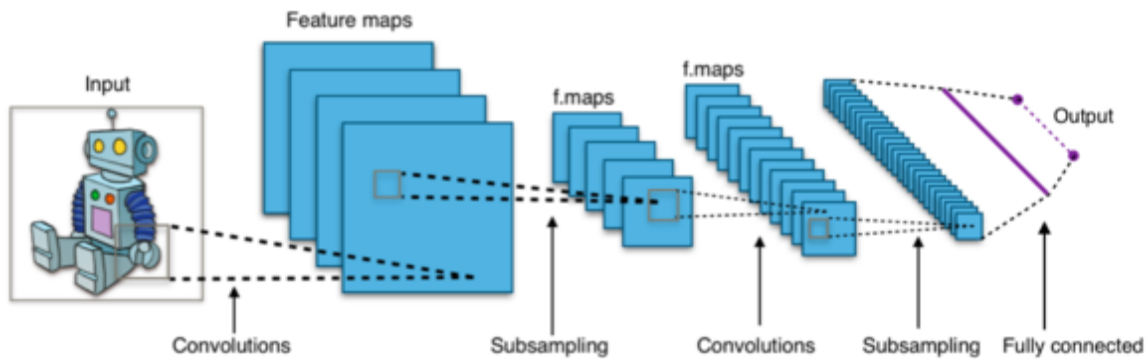

*Figure 1 Typical Convolutional Neural Network Architecture*

Another advantage of CNNs is that they can be trained with relatively small amounts of labeled data. This is particularly useful in the field of image classification, where obtaining large amounts of labeled data can be challenging.

However, CNNs do have some limitations. They require large amounts of computational resources, which can make them difficult to implement on resource-constrained systems. They also require large amounts of labeled data, which can be difficult to obtain in some cases.

In general, CNNs are a powerful tool for microscope cell image classification and have been shown to achieve good results in many studies. However, it is important to underline the fact that they require large amounts of computational resources and labeled data to evaluate their performance using metrics such as accuracy, precision, recall and F1-score.

## Support Vector Machines

Support Vector Machines (SVMs) are a popular method for image classification, including microscope image classification. They have been shown to be effective in many applications, due to their ability to handle high dimensional data and their robustness to overfitting.

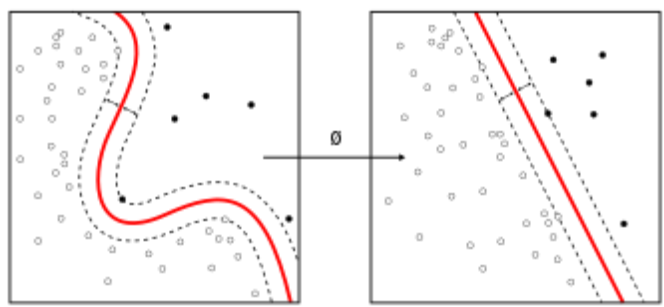

*Figure 2 A linear support vector machine's decision boundary in red (dashed lines showing margins)*

One of the advantages of using SVMs for microscope image classification is their ability to handle non-linearly separable data. SVMs use a technique called the kernel trick to transform the data into a higher dimensional space where it becomes linearly separable. This allows SVMs to effectively classify images with complex features.

Additionally, SVMs can handle at run-time raw, high dimensional data such as microscope images, which can have many features. This is because SVMs implicitly only consider a subset of the training data, called support vectors, in making predictions, which reduces the computational complexity of the model. Another beneficial aspect of using SVMs is their robustness to overfitting, which is a common problem in image classification. This is because SVMs use a regularization term, called the C parameter, to control the trade-off between maximizing the margin and minimizing the classification error.

However, SVMs are not well-suited for large raw datasets, as the training time can be very long. Additionally, SVMs can be sensitive to the choice of kernel function, which requires some experimentation to find the best kernel for a particular dataset. These kinds of limitations need to be considered when applied to real-world problems.

## K-Nearest Neighbors

K-Nearest Neighbors (KNNs) is an old, simple and non-parametric algorithm that can be used for classification. As the name implies, the algorithm assigns class labels to new data points based on the majority class of the data point's k-nearest neighbors.

One of the main advantages of KNN, is that it is an easy-to-implement algorithm (only requires programming the definition of a distance metric) and it is straight-forward to understand and interpret. It also does not require to make assumptions about the data distribution. Another advantage is that KNN is usually robust against noisy data and can be used in multi-class classification problems.

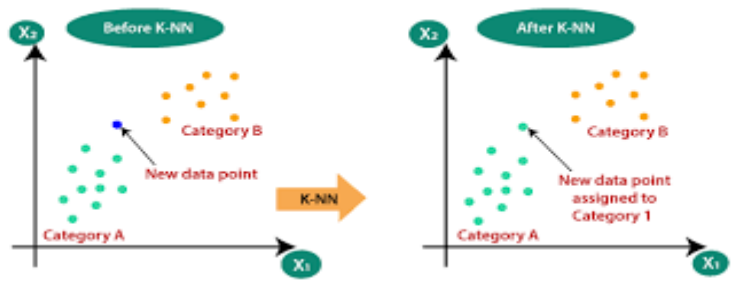

*Figure 3 K-Nearest Neighbors classification*

However, KNN also has several limitations. It is very sensitive to the choice of the value of k, as a large value of k can smooth out noise but can also reduce the decision boundary. It is also very sensitive to the distance metric used, which can affect the performance of the algorithm. Furthermore, it can be proven relatively slow for large datasets and demand high memory requirements.

## Random Forests

Random Forest is an ensemble learning method for classification and regression tasks that can be used for classification. It consists of a collection of decision trees, where each tree is trained on a random subset of the data features.

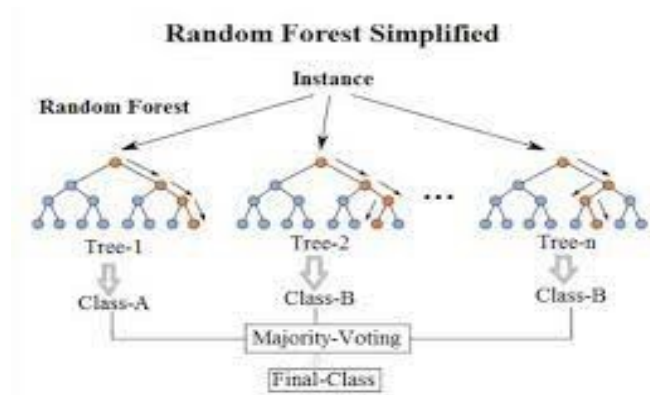

*Figure 4 Random Forest algorithm example*

One of the main advantages of Random Forest is that it is relatively robust to overfitting, as the averaging of the predictions of multiple decision trees can reduce the variance of the final model. It can also handle large amounts of data and it is able to deal with categorical and numerical variables. In addition, Random Forest is an interpretable algorithm, easily understandable that can also provide feature importance measures.

Nevertheless, Random Forest algorithm has some of his own limitations. It can be sensitive to the number of trees used, as a larger number of trees can increase the variance of the final model. The same issue can occur to the number of features used, as a large number of them can increase the correlation between the decision trees. Finally, it is an algorithm that can be highly affected by a class imbalance making it biased towards the majority class.

## Transfer Learning

"Transfer learning" is a technique by which a model trained on one task is used as a starting point for a model on a second, related task. In the context of image cell binary classification, this amounts to using a pre-trained model, such as convolutional neural network (CNN), as a starting point for a new model that will classify images of cells as either infected or healthy.

Transfer learning can be particularly useful in this context because it allows the model to leverage the knowledge learned from the original task (e.g., image classification on large-scale datasets) to improve performance on the new task. This can be especially beneficial when there is limited data available for the new task, as the pre-trained model can provide a good starting point for the new model to learn from.

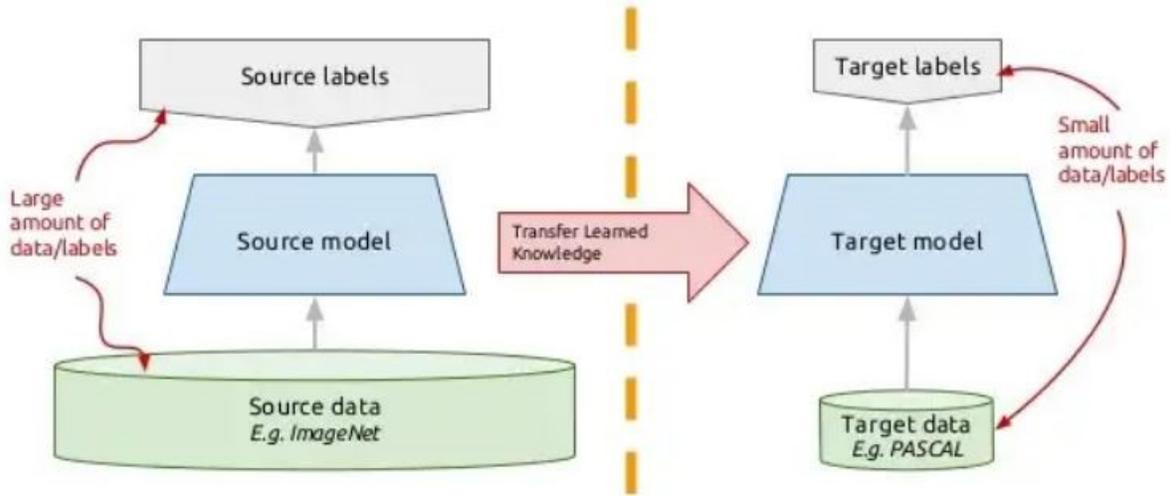

*Figure 5 Example of Transfer Learning technique*

However, it is important to note that not all pre-trained models may be appropriate for a given task, and some fine-tuning of the model may prove necessary. Additionally, while transfer learning can help improve performance, it is not a guarantee, and other techniques as data augmentation may also need to be used.

Overall, Transfer learning is a powerful technique that can be used to improve performance in image cell binary classification tasks with limited data, but it should be used in conjunction with other techniques and careful consideration of the suitability of the pre-trained model. Most known pre-trained models used for the task on hand include Oxfords VGG-16, Resnet-50, Google Inception and MobilenetV2.

For our research study we tried utilizing the pre-trained weights of Resnet-50 and MobileNetV2 models which are pre-trained on the ImageNet dataset by installing them on our code and then we went on adding some additional layers at these pre-trained models such as convolutional layers and fully connected layers and a SoftMax activation function.

## Literature Review

The ability of machine learning to assess and interpret data from numerous diagnostic entities, estimate mortality, and even prescribe suitable treatment based on the patient's characteristics and history makes it a promising tool for the prediction and diagnosis of hematopoietic malignancies.

In (Eckardt et al., 2020) the authors use data mining techniques to predict leukemia by looking at correlations between blood parameters and leukemia, age, and patient health state. Support vector machines (SVM), decision trees (DT), and K-nearest neighbors (CNNs) were employed for prediction.

From a hospital in Gaza, some 4000 patient records were gathered. With a performance rate of 77.3%, decision trees outperformed all classifiers. The authors suggested a supervised machine learning approach for leukemia early prediction (Hossain et al., 2021). The expected symptoms and the likelihood that leukemia would finally be discovered were their main concerns. The training and testing portions of the data set used in this investigation were separated. To evaluate the effectiveness of the model, Naive Bayes, K-Nearest Neighbors, Random Forest, Adaboost, etc. were utilized. A decision tree was used to achieve the best outcome, which had a 98% accuracy rate.

A Support Vector Machine (SVM) was employed by the authors (Furey et al., 2000) to identify leukemia in blood cells. Madhukar et al. (2012) employed k-means clustering (KMC)-based segmentation (MacQueen, J., et al., 1967) to separate the leukocyte nuclei using color-based clustering. The segmented images were used to extract a variety of properties, including Shape (area, perimeter, compactness, strength, eccentricity, elongation, and form factor), GLCM (energy, contrast, entropy, correlation), and fractal dimension.

In order to identify leukemia, Joshi et al. (2013) devised a feature extraction strategy (area, circumference, circularity, etc.) after segmenting blood images. The scientists (Hasan et al., 2017) distinguished lymphocyte cells as blast cells from regular white blood cells using the k-nearest neighbor classifier (KNN). For a lymphoblast classification scheme, Mishra et al. (2019) suggested a discrete orthonormal S-transform-based feature extraction method, followed by a hybrid principal component analysis (PCA) and linear discriminant analysis-based feature reduction strategy. Finally, using an AdaBoost-based Random Forest (ADBRF) classifier, the author (Hasan et al., 2020b) categorised these reduced features.

Deep Learning (DL) is today one of the most popular techniques for supervised classification of raw data; see (Raza et al., 2022) for brain tumor identification, but also (Javeed et al., 2021; Al Razib et al., 2022; Javeed et al., 2022; Ahmad et al., 2020) for many other DL applications for raw data. On the basis of transfer learning, Kasani et al. (2020) introduced the ensemble strategy to categorize cancer cells and normal cells. To counteract the mistake during training, they also used the normalizing technique to shift pixel values between 0 and 1. To deal with the issue of imbalanced classes in their data, they employed a variety of data augmentation strategies. The total accuracy of the ensemble model, which was made up of NASNetLarge and VGG15, was 96.5%.

In another study (Wang, C. et al., 2022), the authors created a supervised deep learning technique that just needs slide-level labels to diagnose malignant hematological disorders. By transforming whole-slide image (WSI) patches into low-dimensional feature representations, the approach increases efficiency. An attention-based network then compiles the patch-level features of each WSI into slide-level representations. Based on these slide-level representations, the model makes final diagnostic predictions. The authors discovered that an area under the receiver operating characteristic curve of 0.966 on an independent test set may be attained at 10 magnifications by applying the suggested model to our assortment of bone marrow WSIs at various magnifications. Additionally, on two publicly accessible datasets, the performance on microscope pictures obtained an average accuracy of 94.2%.

In (K.K. Anilkumar et al.,2022), the authors described the use of DNNs to classify ALL without the need of feature extraction and image segmentation methods. To the best of the authors' knowledge, there is no research on the classification of ALL into subtypes in accordance with the WHO classification scheme using image processing techniques. Future research can seek to detect different forms of leukemic pictures since the current study only evaluated the classification of ALL. The study successfully separated the B-cell and T-cell ALL pictures using a pretrained CNN AlexNet as well as LeukNet, a specially created deep learning network developed by the proposed work, with a classification accuracy of 94.12%. Three alternative training algorithms were used in the study, and the classification performances were also compared.

Modern VGG16 architecture was proposed by (Zakir Ullah et al., 2021) to identify healthy and blast cells in blood smear images. To learn the semantic characteristics, they combined the VGG16 with the Efficient Channel Attention (ECA) module, concentrating on the informative area of the image. This included data magnification, image resizing, and data normalization, among other image preparation techniques. The total accuracy of the VGG16 + ECA model was 91%.

Computers can now read FFR values directly from coronary pictures obtained during CTT angiography thanks to a ground-breaking deep neural network technique (TreeVes-Net) put forth by researchers (Xu et al., 2021). Their solution uses a tree-structured recurrent neural network to record data relating to blood fluid and coronary channel geometry (RNN). They obtained 0.92 and 0.93 in the area under the ROC curve AUC in experiments using 180 real trees from clinical patient data and 13,000 fake coronary trees, respectively.

The VIT-CNN ensemble strategy, which combines the EfficientNetB0 with the vision transfer model for b-lymphoblast detection, was first described by Jiang et al. in 2021. To prevent overfitting, they normalize and transform the image's size. To increase the number of photographs in the dataset, they employ a novel method to enhance data selection. In the test set, the VIT-CNN ensemble network had a 99.03% accuracy rate.

A hybrid approach combining a CNN-based VGGNet model and the Salp Swarm algorithm was presented in (Sahlol et al., 2020). In this hybrid approach, called SASSA, features were extracted using a pre-trained VGG-net model, and SASSA was used to select significant features, reduce noisy features, and boost the model's accuracy. Cells were categorized as normal or pathological using SVM. For the ALL -IDB2 dataset, the SVM classifier had an accuracy of 96.11%, and for the ISBI-2019 dataset, it had an accuracy of 87.9%.

UNET- and UNET++-based approaches for segmenting leukocyte nuclei have been developed recently to improve the categorization of normal and blast cells (Jiang et al. 2021; Alagu et al., 2021).

The AlexNet model was suggested by the authors (Shafique et al., 2018; Loey et al., 2020) for the recognition of ALL from microscopic blood smear pictures based on transfer learning. In order to deal with the issue of insufficient data, data enrichment techniques are also discussed.

To categorize the normal and blast cells from the ISBI-2019 dataset, Ghaderzadeh et al. (2022) developed a majority voting ensemble technique incorporating the four models (InceptionV3, ResNet-V2, Xception, DesNet121). The ensemble model obtained 98.5% accuracy following preprocessing and expansion. Two HISTOCNN and HistoNet models based on CNN architectures were presented by (Genovese et al., 2021).

The HistoNet model used the ALL-IDB dataset to classify normal and blast cells by adopting the features of the HistoCNN model based on transfer learning.

The nucleus of WBC was segmented using K-means clustering, C-means, marker-controlled watershed, and histogram-based thresholding approaches (Gebremesket et al.,2021; Kandhari et al.,2021). Based on microscopic blood smear images, the scientists offered both individual and ensemble models for the detection of ALL cancer, although the ensemble models outperformed the individual models in terms of accuracy. The weighted ensemble of the network model (Chen et al., 2021) obtained 88.3% accuracy, whereas the ensemble model ResNet101-9 (Bodzaas et al., 2020) achieved 85.11% accuracy.

## Dataset

The main dataset that was used for our study was the C-NMC 2019 dataset which is publicly available. The source to our dataset was the [Cancer Image Archive (TCIA),](#) a service which de-identifies and hosts a large archive of medical images of cancer, accessible for public download. TCIA is funded by the Cancer Imaging Program (CIP), a part of the United States National Cancer Institute (NCI) and is managed by the Frederick National Laboratory for Cancer Research (FNLCR).

The data are organized into collections; typically, patient images related by a common disease (e.g., lung cancer), image modality or type (MRI, CT, digital histopathology, etc.) or research focus. DICOM is the primary file format used by TCIA for radiology imaging. An emphasis is made to provide supporting data related to the images such as patient outcomes, treatment details, genomics and expert analyses. To enhance the value of TCIA's collections it is encouraged by the research community to publish their analysis results.

TCIA in fact is a large public archive of medical images and related data focusing on cancer imaging. It is a resource that provides access to a wide range of medical imaging data, including images from computed tomography (CT), magnetic resonance imaging (MRI), and position emission tomography (PET), as well as clinical and demographic information.

One of the main advantages of TCIA is the large size and diversity of data. The archive contains over 30.000 patients and over two million imaging studies, covering a wide range of cancer types and modalities. This allows researchers to access a large and diverse dataset for developing and evaluating computational models and algorithms for image analysis and diagnosis.

Another advantage of TCIA is the ease of use. The archive is freely accessible through a web-based interface, which allows users to easily search, view, and download the data. The collections are also well-organized and annotated which makes it easy to find and use the data you need.

Furthermore, TCIA provides a wide range of tools and resources to support the use of data. These include software tools for image viewing and analysis, as well as a variety of tutorial, guides, and other resources to help researchers get started with the data.

In summary, TCIA is a valuable resource for researchers in the field of medical image analysis and cancer research. Its large size and diversity of data and wind range of tolls and resources make it an ideal resource for developing and evaluating computational models and algorithms for image analysis and diagnosis.

### C-NMC Dataset

The Dataset used for the purposes of this study was the C_NMC_2019 dataset, downloaded from

https://wiki.cancerimagingarchive.net/pages/viewpage.action?pageId=52758223 .

This dataset aims to provide a source for image classification of leukemic B-lymphoblast cells from normal B-lymphoid precursors (normal cells) from blood smear microscopic images.

A dataset of cells with labels (normal versus cancer) is provided to train machine learning-based Classifiers, in order to identify normal cells from leukemia blasts. These cells have been segmented from the microscopic images. The images are representative of images in the real-world because these contain some staining and illumination errors, although these errors have largely been fixed on the original set using stain color normalization.

This dataset is a valuable resource for researchers in the field of biomedical engineering and machine learning as it provides a large set of cell images to train and test algorithms on.

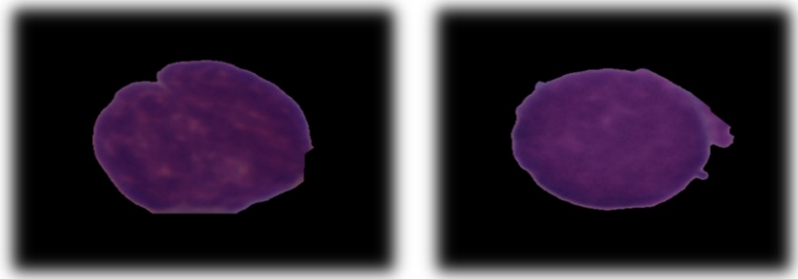

Figure 6 ALL-Leukemia image cell (Left) versus Healthy image cell (Right)

As shown in Figure 6, healthy cells and cancer cells can appear similar under certain imaging techniques, such as microscopy. This can make it difficult for human observers to accurately identify cancerous cells. Machine learning algorithms can be trained to recognize patterns in images that may be difficult for humans to discern, making it a useful tool for identifying cancer cells in images. Additionally, machine learning can be used to classify images automatically and with high accuracy, which is crucial in the medical field where time is often of the essence. Blood cells appear purple in images because they contain a protein called hemoglobin, which is responsible for carrying oxygen throughout the body. Hemoglobin is a reddish-brown color when it is carrying oxygen, but when it is not carrying oxygen, it appears purple. When blood cells are viewed under a microscope, the purple color of the deoxygenated hemoglobin can be seen. Additionally, the purple color may also be due to the staining techniques used during preparation of the blood sample for imaging. The purple color of the blood cells in the images is usually due to the combination of the hemoglobin color and the staining agents used.

The dataset contains RGB image cells that have a resolution of 450x450 and are derived from a total of 118 individual subjects, distributed as follows:

- ALL (cancer) subjects: 69
- Normal subjects: 49

**Train set composition:**

- Total subjects: 73, ALL: 47, Normal: 26
- Total cells: 10,661, ALL: 7272, Normal: 3389

**Preliminary test set composition**:

- Total subjects: 28, ALL: 13, Normal: 15
- Total Cells: 1867, ALL: 1219, HEM: 648

**Final test set composition:**

- Total subjects: 17, ALL: 9, Normal: 8
- Total Cells: 2586

It is important to notice from the data distribution above that the true labels of the final test were not provided. The purpose of this was that the results of the classification should be checked at the leaderboard of the codalab challenge to know the comparative results with the world teams. As a result, we did not have label information on the test data, so to evaluate the performance of our approach, we resorted to the classical train/test split of the labeled data we had in our disposal as shown in Table 1 below.

*Table 1 Overview of C-NMC dataset*

|                     | Number of Cell Images | ALL cells (leukemia) | Healthy Cells |
| ------------------- | --------------------- | -------------------- | ------------- |
| Dataset             | 10.661                | 7272                 | 3389          |
| Training Set (80%)  | 8528                  | 5817                 | 2711          |
| Test Set (20%)      | 2133                  | 1455                 | 678           |

As present in Table 1, we split the original labeled data in training and test set. A commonly used ratio for dividing a dataset into training and test set is 80/20 or 70/30, where 80% or 70% of the data is used for training and the rest 20% or 30% is used for testing. However, the appropriate ratio can vary depending on the specific application and the size of the dataset. In some cases, it may be beneficial to use a smaller test set, such as a 50/50 split, to ensure that the model is tested on a diverse set of examples.

# Proposed Methodology

In this section we describe the steps we followed in order to build a machine learning pipeline that ultimately provides good results in the binary image classification on whether the cell is healthy or drops to the cancer (ALL) category.

After collecting the data, these steps include data pre-processing, analysis of model architectures available, choosing architectures that were more fitting to our problem, testing them on the pre-processed dataset and then after obtaining results we tried to tune the hyperparameters of our models in order to achieve an even better outcome.

It is important to highlight at this point, that all the models that we experimented with, are Convolutional Neural Network based models (CNNs). The reason behind this selection was that CNNs are particularly well-suited for cell image classification because they are able to effectively learn and extract feature for images.

Hereby, we present some of the main reasons of our decision:

- Spatial information: CNNs are able to take into account the spatial structure of the images through the use of convolutional layers, which scan the images with a set of filters to detect features such as edges, patterns, and textures.
- Scale invariance: CNNs have the ability to learn features that are invariant to changes in scale, rotation, and translation. This is achieved through the use of pooling layers, which reduce the spatial resolution of the feature maps, but retain important information.
- Transfer learning: CNNs have been trained on large image datasets such as ImageNet, so it is possible to use pre-trained models and fine-tune them to specific datasets. This allows the model to take advantage of the knowledge learned from other datasets to improve the performance on the specific task.
- Automated feature learning: CNNs can learn useful features directly from the data, without requiring any manual feature engineering. This is particularly useful for cell image classification, where the features are not easily defined.
- High performance: CNNs have been shown to achieve state-of-the-art performance on a wide range of image classification tasks, including medical image classification.

In summary, CNNs are able to handle the complexity and variety of cell images and extract informative features that are useful for classification. They are able to learn and generalize well to unseen data, and they have been proven to achieve high performance on cell image classification tasks which concluded in choosing them as the main base of the model architectures we chose to use.

## Data Pre-Processing

Data pre-processing is a crucial step in image classification, including binary classification of cell images. This is because the quality of the input data can greatly affect the performance of the classifier.

In our study we present some of the crucial steps of our internal data pre-processing pipeline that played a vital role in increasing our model performance and removing some image features that our classifier would incorrectly learn from and be biased to.

## Image Resizing

Image resizing is an important step in pre-processing for cell images classification. The reason for this is that the classifier needs the images to be of consistent size and aspect ratio in order to learn generalizable features. So, in order to change the image size, we must ensure features like the aspect ratio are preserved.

A larger image size typically means more pixels, which means more data for the classifier to process. This can lead to the classifier taking longer to run because it has to process more information. Additionally, larger images may also require more memory, which can also slow down the classifier

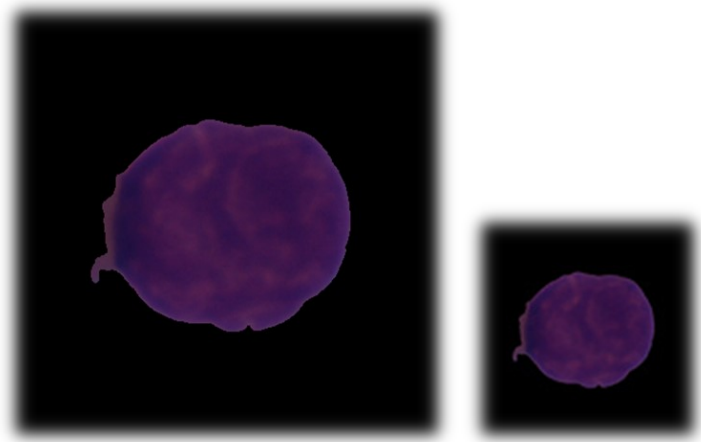

*Figure 7 Image Resizing from size 450X450 to 128X128*

When the images are of different sizes, the classifier may struggle to learn features that are common to all the images, which can negatively impact its performance. Additionally, some classifiers, such as convolutional neural networks (CNNs), have a fixed input size, and require images to be resized to that size.

Resizing the images (see fig. 7) can be done using different methods. Interpolation methods such as bilinear, bicubic, and nearest-neighbor interpolation can be used to resize the images. Bilinear interpolation is generally considered to be the best choice, as it provides a good balance between preserving image quality and computational cost. Images can be resized while preserving their aspect ratio by using the aspect ratio of the original image. Resizing an image can affect cell image classification in several ways:

- Interpolation: Depending on the method used to resize the image, the quality of the image can be degraded. For example, if the image is resized using simple interpolation methods such as nearest neighbor or bilinear interpolation, the resulting image may contain artifacts such as jagged edges or blurriness.
- Aspect ratio: If the aspect ratio of the image is not preserved during resizing, the shape of the cells may be distorted. This can make it difficult for a classifier to accurately identify the cells.

- Scale: Resizing the image can also change the scale of the cells, which can make it difficult for a classifier to distinguish between small and large cells.
- Resolution: The resolution of the image can also be affected by resizing, which can make it difficult for a classifier to detect small details in the image.

It's important to note that in some cases, resizing the image can improve the performance of the classifier, for example if the image is too big for the classifier to handle, or the cells are too small in the original image, but it's important to do it carefully and with the appropriate method to preserve the image quality and not distorting the cells. For our case of study, we tried many different image resizing patterns. In overall, instead of the original 450x450 image size that our dataset had, we experimented on image sizes of 64x64, 128x128 and 256x256 as well as 224x224 as that was the size of the original training image for some pre-trained deep learning algorithms that we experimented on using transfer learning.

## Image Rotation

A classifier can be biased towards a specific image direction if the training data primarily contains images that are positioned in that specific direction. For example, if the majority of the images in the training data are rotated towards a specific orientation, then the model will learn to recognize objects primarily in that orientation. As a result, the model may have difficulty recognizing the same object in other orientations, leading to poor performance on images that are rotated in a different direction.

This bias can be caused by a number of factors such as data collection process, data labeling, and data preprocessing. For example, if the data collection process is biased towards capturing images in a specific orientation, or if the data labeling process only includes images that are rotated in a specific direction, this can lead to a bias in the training data. Similarly, if the data preprocessing step only includes rotating the images in a specific direction, this can also lead to a bias in the training data.

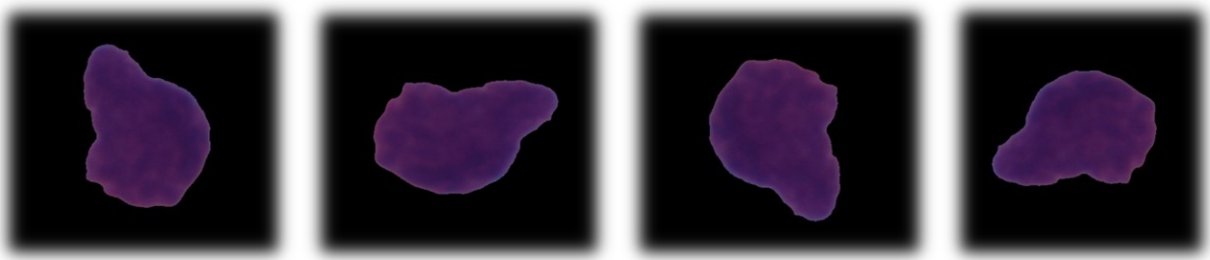

*Figure 8 Image rotation example on 4 angles (0°,90°,180°,270°)*

To mitigate this problem, it is important to use a diverse set of training images that are rotated in various orientations, and use data augmentation techniques, like random rotation, to increase the diversity of the training data.

Image rotation (see fig. 8) is an important pre-processing step in image classification because it allows the model to learn and recognize objects in the image regardless of their orientation. This is important because in real-world scenarios, an object may be captured at various orientations and the model should be able to correctly classify it regardless of the orientation. By rotating the images during training, the model will be exposed to different orientations of the same object, improving its ability to generalize to new, unseen images. Additionally, rotating the images during training also helps to increase the diversity of the training data, which can lead to better performance. In general, rotating the images clockwise by 90, 180, and 270 degrees will cover all 4 angles of the image. This is a common approach when the goal is to ensure that the model can recognize the object in the image regardless of its orientation and this was the approach we followed in our study. However, if the goal is to only recognize the object in one specific orientation, then rotating the image by only that specific angle will be sufficient. Additionally, it is also possible to use data augmentation techniques to randomly rotate the images during training, which can help to further increase the diversity of the training data and improve the model's ability to generalize to new, unseen images.

As shown in Table 2, our training dataset was 4 times larger than the original after adding copies of the original images rotated by 90,180 and 270 degrees.

*Table 2 Overview of C-NMC dataset after adding rotated samples to the training set*

|  | Number of Cell Images | ALL cells (leukemia) | Healthy Cells |
|---|---|---|---|
| Dataset | 42644 | 29088 | 13556 |
| Training Set (80%) | 40511 | 27633 | 12878 |
| Test Set (20%) | 2133 | 1455 | 678 |

## Grayscale Transformation

Grayscale transformation can help in image classification by reducing the dimensionality of the data and removing redundant information. In a grayscale image, each pixel is represented by a single value, whereas in a color image each pixel is represented by three values (one for each of the red, green, and blue channels). This can make it easier for a classifier to learn the features that are important for making a prediction, as well as reducing the number of computational resources required to process the image. Additionally, grayscale images are less sensitive to variations in lighting conditions and can be more robust to changes in the environment.

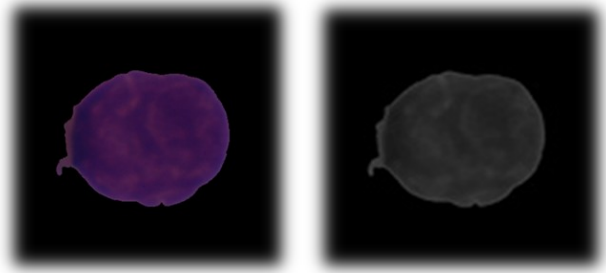

*Figure 9 Example of Grayscale Transformation on our Dataset*

In addition, grayscale transformation (Figure 9), can make it easier for a classifier to identify the shape, size, and texture of the cells, which are important features for cell classification. Furthermore, since grayscale images have less noise and less variation in color, the classifier may be less likely to be thrown off by small variations in lighting or other factors. Furthermore, grayscale images tend to be smaller in size, which allows the classifier to run faster on the image dataset and can be beneficial in large image datasets.

Another example of how grayscale transformation can be beneficial to cell image classification is to use grayscale image as input to convolutional neural network (CNN) to classify cells which is applicable in our study. In this case, grayscale images can be helpful in reducing the computational complexity and memory requirements of the CNN, while still allowing the network to effectively learn relevant features for cell classification.

In addition, grayscale transformation can be used to enhance the contrast in the images, making it easier to identify the cells and their features. This can be particularly useful when working with low-quality images or images with poor lighting conditions.

Finally, grayscale transformation is a good starting point in image classification in machine learning because it simplifies the input data by converting the image from a 3-channel RGB image to a 1-channel grayscale image.

## Image Normalization

Image normalization is important in cell image classification because it helps to improve the performance and stability of the classifier by ensuring that the input images have similar scales and contrasts. This is particularly important when using convolutional neural networks (CNNs) because they are sensitive to the scale and contrast of the input images. Normalization can be achieved by transforming the image pixel values so that they have a mean of zero and a standard deviation of one, or by rescaling the pixel values to a fixed range, such as [0, 1]. By normalizing the input images, the CNN can focus on learning the features and patterns in the images rather than being influenced by variations in scale and contrast.

Convolutional Neural Networks (CNNs) benefit from image normalization before classification because normalization helps to improve the stability and performance of the CNN by ensuring that the input images have similar scales and contrasts. CNNs are designed to learn features and patterns in images by applying a series of convolutional and pooling layers to the input images. These layers are designed to extract features from the images by applying a set of filters to the input images. If the input images have different scales and contrasts, the filters may not be able to extract the same features from all images, which can affect the performance of the CNN.

By normalizing the input images, the CNN can focus on learning the features and patterns in the images rather than being influenced by variations in scale and contrast. This can help to improve the stability and performance of the CNN by reducing the impact of variations in scale and contrast on the output of the CNN. Additionally, it can help to reduce the chances of overfitting by making sure that the input image is in a standard format, and also can reduce the training time by reducing the number of parameters that the CNN needs to learn.

## Image Flipping

Flipping an image (different from image rotation, see fig. 10) is good when training a CNN model for image classification because it increases the diversity of the training data and makes the model more robust to different variations of the input images. When a CNN model is trained on a limited set of images, it may only learn to recognize certain features and patterns that are present in those images. By flipping the images (around a chosen axis), it can learn to recognize the same features and patterns from images taken from different view-points. This can make the model more robust to different variations of the input images and improve its performance on unseen data.

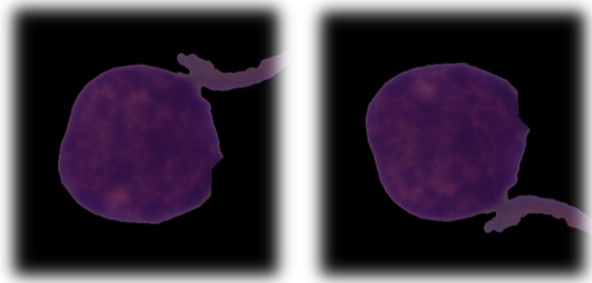

*Figure 10 Example of flipping a cell image in the dataset*

Additionally, flipping an image can also introduce new variations to the training data, such as changes in the relative position and orientation of the objects in the image, as well as changes in lighting and shadows. These variations can help the model to generalize better and improve its performance on unseen data. Furthermore, flipping can also be used as a data augmentation technique to artificially inflate the size of the dataset, which can help to reduce overfitting and improve the model's performance.

In summary, flipping an image is a good practice when training a CNN model for image classification because it increases the diversity of the training data, makes the model more robust to different variations of the input images, and can help to reduce overfitting and improve the model's performance.

Image flipping and image rotation are both types of data augmentation techniques used in image classification tasks. Data augmentation is a technique used to artificially increase the size of a dataset by applying various transformations to the existing data. This helps to improve the performance and robustness of machine learning models, as they are exposed to a wider variety of data during training.

To summarize, it is important to note that data augmentation should not be performed on the validation set. The validation set is used to evaluate the performance of a model and determine its generalization ability, so it should reflect the distribution of the unseen data as closely as possible. Data augmentation is typically used to artificially increase the size of the training set and make the model more robust to variations in the input data, but it should not be used on the validation set as it would artificially inflate the performance of the model.

## Proposed Model Architecture

A default CNN model typically consists of several convolutional layers, interspersed with non-linear activation functions such as ReLU, and followed by one or more fully connected layers. The convolutional layers are responsible for extracting features from the input image, while the fully connected layers classify the image based on those features. The model is trained using a dataset of labeled images, and the objective is to minimize the difference between the predicted labels and the true labels.

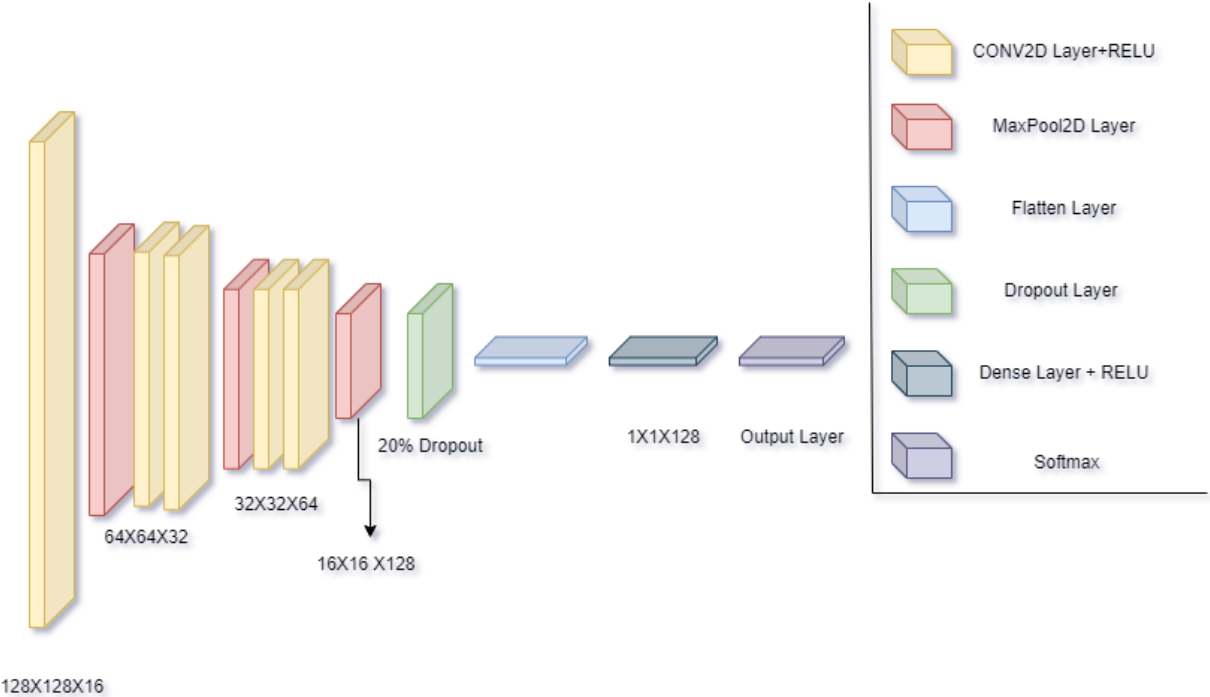

*Figure 11 Structure of our Proposed CNN model*

The architecture of our CNN model shown in fig. 11, consists of five convolutional layers, each followed by a max pooling layer to reduce the spatial dimensions of the feature maps. The convolutional layers use a kernel size of 3x3 and a stride of 2, and are designed to learn 16, 32, and 64 filters respectively. The convolutional layers are followed by a dropout layer with a rate of 0.2 to reduce overfitting.

After the convolutional layers and the dropout, we use a flatten layer to convert the feature maps into a 1-dimension vector that serves as the input to a fully connected layer with 128 neurons, followed by a ReLU activation function. The final fully connected layer has an output of 2 neurons followed by the Softmax activation function, corresponding to the binary classes in our dataset. The model is trained using Sparse Categorical Cross-entropy, and an RMSprop optimizer with a learning rate of 0.0001.

## Convolutional Layers

The number of convolutional layers in a CNN is an important design decision that can greatly impact the model's performance. In our model, we have used five convolutional layers. The reason for choosing this number of layers is that it provides a good balance between model complexity and performance.

The first three layers were designed to learn simple features such as edges, lines and corners, which are important for recognizing basic shapes and patterns in the input images. The third and fourth layers were designed to learn more complex features such as textures and patterns, which are important for recognizing more detailed features in the input images.

We have used five layers based on the observation that more layers allow the model to learn more abstract and high-level features that are useful for recognizing objects in the input images. Additionally, we have used five layers to increase the model's capacity to learn complex relationships between the input and output. We have also tried different number of layers and found that five layers gave us the best results in terms of accuracy and generalization.

In conclusion, the number of convolutional layers in our model has been chosen based on a trade-off between model complexity and performance, also considering the computational cost and the overfitting risk.

## Kernel Size

In our model, we have used a kernel size of 3x3 for all the convolutional layers. The kernel size is the size of the filter that is applied to the input image in the convolutional layer. The filter is a matrix of weights that is used to extract features from the input image. A larger kernel size allows the model to extract larger features from the input image, while a smaller kernel size allows the model to extract smaller features.

We have chosen a kernel size of 3x3 because it is small enough to extract small features such as edges, lines and corners, but also large enough to extract larger features such as textures and patterns. This is important for recognizing basic shapes and patterns in the input images, and also more detailed features in the input images.

We also tried different kernel sizes such as 5x5 and 7x7 but found that a kernel size of 3x3 gave us the best results in terms of accuracy and generalization. A larger kernel size increases the model's capacity to learn complex relationships between the input and output, but it also increases the number of parameters and computation cost, which can lead to overfitting and must be taken in consideration given the hardware equipment limitations.

## Max pooling Layers

A max pooling layer (fig. 12) is a type of pooling layer that is commonly used in Convolutional Neural Networks (CNNs). The purpose of a max pooling layer is to down-sample the spatial dimensions of the feature maps generated by the convolutional layers. This is done by applying a max pooling operation to small non-overlapping regions of the feature maps.

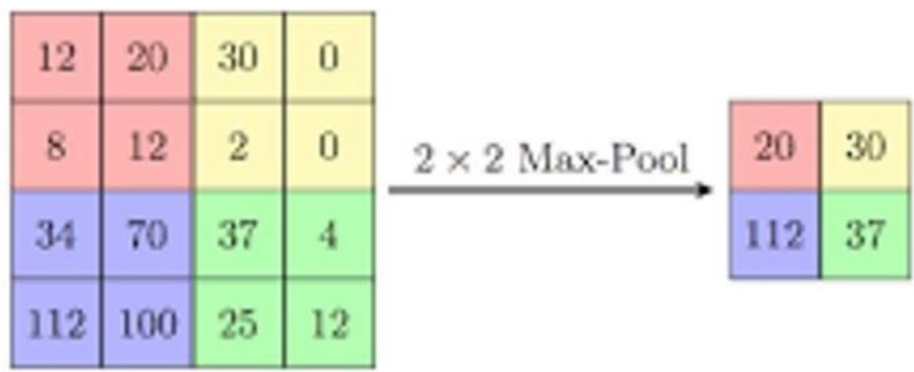

*Figure 12 Example of down-sampling using Max-Pooling*

The max pooling operation works by taking the maximum value from each region, and outputting that value as the corresponding element in the down-sampled feature map. This has the effect of keeping only the most important information from each region, while discarding less important information. The main reason for putting a max pooling layer after a convolutional layer is to down-sample the spatial dimensions of the feature maps, which can be useful for reducing computational cost, increasing the model's ability to generalize and decreasing overfitting risk, while maintaining important information.

Strides are commonly used in max-pooling layers in order to down-sample the feature maps. Using strides in max-pooling layers can help to reduce the size of the feature maps, which can be useful when working with large images or when the computational resources are limited. A stride of 2 for example will reduce the feature maps by half in each dimension.

Using strides in max-pooling layers can also help to reduce the number of parameters in the model, which can be useful when working with a limited amount of training data. Another prospect from using strides in max-pooling layers, is that it can also help to improve the model's ability to generalize to new data, as the model is forced to learn more abstract and high-level features that are useful for recognizing objects in the input images.

On the other hand, not using strides in max-pooling layers can help to preserve the spatial dimensions of the feature maps, which can be useful when working with images that have a consistent border or when the spatial information is important for the task.

## Dropout Layer

A dropout layer is a regularization technique used in convolutional neural networks (CNNs) to prevent overfitting. Overfitting occurs when a model is trained too well on the training data and performs poorly on unseen data.

Dropout works by randomly dropping out (i.e., setting to zero) a certain percentage of the neurons during each training iteration. This forces the remaining neurons to learn more robust features, since they can no longer rely on the dropped-out neurons. This can help to prevent overfitting by making the model more robust to changes in the input data.

In a CNN, dropout is typically applied after the convolutional layers and before the fully connected layers. This is because the convolutional layers are responsible for extracting features from the input image, and the fully connected layers are responsible for making the final classification decision. By applying dropout after the convolutional layers, the model is forced to learn more robust features that are less likely to overfit. We added a dropout layer with a 20% percentage of dropout after our convolutional layer blocks and before the our fully connected layer.

## Flatten Layer

A flatten layer is an important component in a convolutional neural network (CNN) model for binary cell image classification because it is responsible for reshaping the output of the previous layers from a multi-dimensional tensor to a one-dimensional tensor. This is necessary because the output of the previous layers represents the features that have been extracted from the input image, and these features need to be presented in a format that can be used by the fully connected layers that follow.

The flatten layer takes the output of the previous layers, which is typically a multi-dimensional tensor with dimensions such as width, height, and depth, and reshapes it into a one-dimensional tensor. This one-dimensional tensor can then be used as input to the fully connected layers, which are responsible for making the final classification decision.

The reason why a flatten layer is important in binary cell image classification is because it allows the model to make the final classification decision based on the features that have been extracted from the input image. This is important in binary cell image classification because the model needs to be able to distinguish between cells of different classes based on their characteristics, such as shape, size, and texture.

## Activation Functions

In our experiments we utilized a couple of activation functions such as ReLU (fig. 13), Softmax and Sigmoid. Eventually we decided to use the ReLU activation function, commonly used in a convolutional neural network (CNN), specifically in the convolutional layers and fully connected layers, and the Softmax activation function in the output layer.

Rectified Linear Unit (ReLU) activation function is commonly used in binary image cell classification because it has several benefits that make it well-suited for this task. ReLU introduces non-linearity into the model, which allows the model to learn more complex and abstract features from the input image. This is important in binary image cell classification because the model needs to be able to distinguish between cells of different classes based on their characteristics, such as shape, size, and texture. Another benefit of using ReLU as an activation function in CNNs for binary image cell classification is that it helps to alleviate the vanishing gradient problem. This problem occurs when the gradients of the weights become very small during training, making it difficult for the weights to be updated. The ReLU activation function helps to alleviate this problem by introducing non-linearity into the model, which allows the gradients to be updated more easily.

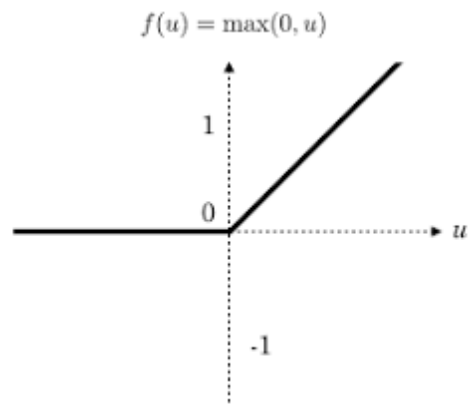

*Figure 13 ReLU Activation Function*

In addition, ReLU is computationally efficient as it does not require any complex mathematical operations, such as exponential or logarithmic functions. This can be beneficial for CNNs for binary image cell classification as it can reduce the amount of computation required during training, which can make the training process faster and more efficient.

In order to choose the activation function for our output layer we took in consideration that in binary image classification, the goal is to predict one of two possible labels. Softmax is a commonly used activation function in the output layer of a neural network for this task because it allows the model to output a probability distribution over the two classes, rather than just a single output value. This allows the model to express its level of confidence in its prediction, which can be useful in certain applications.
Additionally, using softmax as the output activation function allows the network to be trained using cross-entropy loss, which is a common loss function for multi-class classification problems.

## Optimization Algorithm
When compiling our model, we experimented with Adam solver and RMSprop as choices for the optimization algorithm. RMSprop and Adam are both popular optimization algorithms that are commonly used in image classification tasks. Finally, after experimenting on both with different kinds of learning rates, we evaluated that RMSprop optimizer provides better results towards the prediction accuracy of our classifier.

RMSprop is known for its ability to adaptively adjust the learning rate for each parameter in the model. This allows the algorithm to converge faster and with better stability, which is particularly important for image classification tasks where the data can be highly non-linear and complex. Additionally, RMSprop uses a moving average of the squared gradient values which helps to dampen the oscillations that can occur during the optimization process.

## Transfer Learning Implementations

Aside from our proposed CNN model, in this study we experimented on transfer learning techniques that allows researchers to use pre-trained models on large datasets. The two models that we used as a base to our transfer learning architectures were ResNet50 and MobilenetV2. Both of those models were pre-trained using the ImageNet dataset.

ImageNet is a large dataset of labeled images that is widely used in the field of computer vision. It was first introduced in 2009 by researchers at Stanford University and has since become a standard benchmark for image classification and object recognition tasks. The dataset contains over 14 million images, with a total of over 22,000 categories, including both general and fine-grained classes. The images are sourced from the Internet and have been labeled by human annotators. The dataset has been used in many influential research papers and has played a significant role in the development of deep learning techniques for image recognition and classification. However, it has also been criticized for its lack of diversity and potential biases in the data and annotations.

The first step in enforcing a transfer learning technique is to load the pre-trained weights of the models available. In our task we used the Tensorflow library to load the pre-trained weights of RESNET-50 and MOBILENETV2 models. Next, we proceeded to remove the last fully connected layer of the pre-trained model as this layer is specific to the original task the model was trained on. Furthermore, we experimented on adding our own fully connected layer corresponding to the number of classes for our binary classification problem. One of the most important tasks of this process was to freeze the layers of the pre-trained model. This is an important step, as we do not want the pre-trained weights to get updated during training. In that manner, we take advantage of the features already learned by the pre-trained model.

Finally, after the utilization of the pre-trained model and the enhancement with extra layers of our choice, we train the output model using our own dataset by compiling it with an optimizer, a loss function and an evaluation metric such as accuracy.

## RESNET-50 Model Architecture

ResNet-50 is a pre-trained convolutional neural network (CNN) architecture that has been trained on the ImageNet dataset, which contains over 14 million images and 1000 classes. The architecture is a deep residual network, where the residual connections allow the network to learn more complex features.

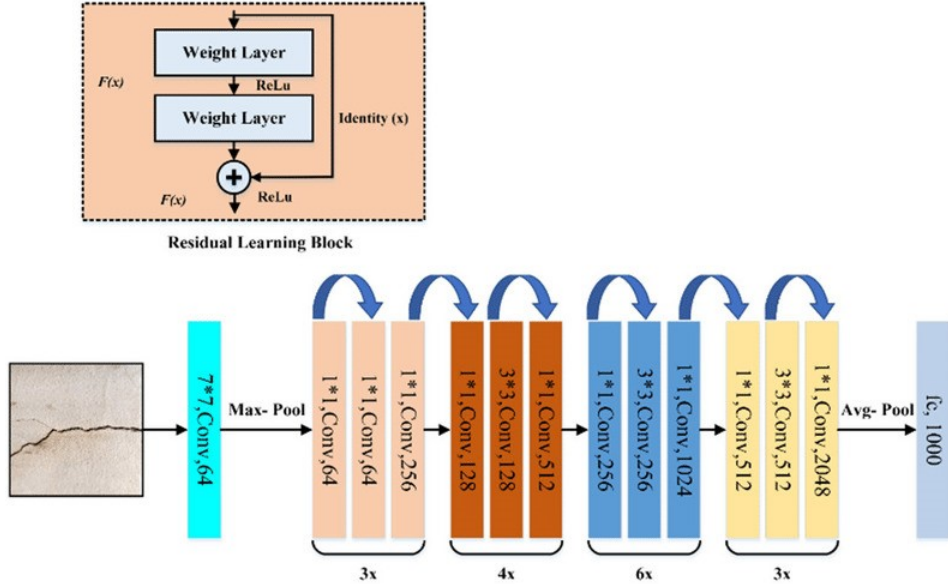

*Figure 14 ResNet-50 Model Architecture*

The ResNet-50 architecture (fig. 14) has 50 layers, including 50 convolutional layers, and the number of filters increases as the depth of the network increases. The architecture also includes batch normalization layers and ReLU activation functions after each convolutional layer. The architecture of the ResNet-50 model is divided to the following stages:

- The first stage has a convolutional layer with a kernel size of 7x7 and a stride of 2, followed by a maximum pooling layer with a kernel size of 3x3 and a stride of 2.
- The second stage is the first convolutional layer block, containing nine convolutional layers overall. These layers follow begin  3×3, 64 kernel convolution, another with 1×1, 64 kernels, and a third with 1×1, 256 kernels. These three layers are repeated three times.
- The third stage is the second convolutional layer block, containing twelve convolutional layers overall. These layers follow begin  1×1, 128 kernel convolution, another with 3×3,128 kernels, and a third with 1×1, 512 kernels. These three layers are iterated four times.
- The fourth stage of the model is the third convolutional layer block, containing eighteen convolutional layers overall. These layers follow begin  1×1, 256 kernel convolution, another with 3×3,256 kernels, and a third with 1×1, 2048 kernels. These three layers are repeated a total of six times.
- The fifth stage of the model, which is the last convolutional layer block, contains nice convolutional layers. These layers follow begin  1×1, 512 kernel convolution, another with 3×3,512 kernels, and a third with 1×1, 2048 kernels. These three layers are repeated three times.

- The final stage has a global average pooling layer that averages the outputs of the third stage across the spatial dimensions, followed by a fully connected layer that produces the final classification output.

ResNet-50 uses skip connections to add the outputs of the lower layer to the outputs of the higher layers, this allows the gradients to flow more easily through the network, which helps to prevent the vanishing gradient problem. The ResNet-50 architecture is particularly well suited for transfer learning, as it has been trained on a large dataset, allowing it to learn robust features that can be useful for other image classification tasks.

In transfer learning, the pre-trained ResNet-50 model can be used as a feature extractor by removing the last fully connected layer, which is specific to the original task the model was trained on and adding a new fully connected layer for your specific task. The other layers of the model can be frozen to preserve the learned features and improve the training performance.

The ResNet-50 model is widely used in a variety of computer vision tasks, such as image classification, object detection, and semantic segmentation.

In summary, ResNet-50 is a pre-trained deep residual architecture model trained on the ImageNet dataset, which contains over 14 million images and 1000 classes. The model's architecture is designed to learn more complex features, which makes it suitable for transfer learning. The model is well suited for image classification tasks and is easy to fine-tune on new datasets.

## MOBILNETV2 Model Architecture

MobileNetV2 is a pre-trained convolutional neural network (CNN) architecture that was designed to be lightweight and efficient for mobile and embedded devices. The architecture is built upon depthwise separable convolutions, which significantly reduces the number of parameters and computation required, while still preserving the accuracy of the model.

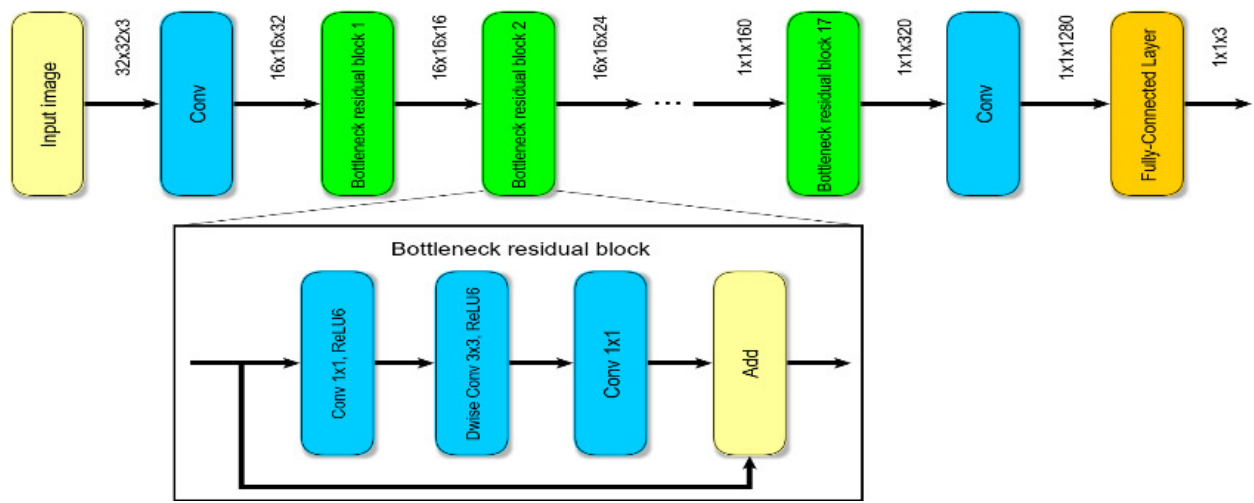

*Figure 15 MobileNetV2 Model Architecture*

The MobileNetV2 architecture (see fig. 15) has 19 layers, including 13 depth wise separable convolutional layers and 6 pointwise convolutional layers, and has a width multiplier that controls the number of filters in each layer. The depth wise convolutions are used to extract spatial features, while the pointwise convolutions are used to increase the number of filters in the network.

The final block of the network contains a global average pooling layer and a fully connected layer for classification. The architecture also includes batch normalization layers and ReLU activation functions after each convolutional layer. The MobileNetV2 architecture is based on the inverted residual structure, where a standard convolution is replaced by a depth wise convolution followed by a pointwise convolution. This allows for a more efficient use of computation resources, as the depth wise convolution is less computationally expensive than a standard convolution. The architecture also uses a bottleneck layer, which reduces the number of filters in the network, further improving computational efficiency.

MobileNetV2 has been widely used in image classification tasks and it has been demonstrated to be an effective model for transfer learning. It's small size and low computational requirements make it suitable for our task given our system specifications which brought limitations to our study.

## Computational Results

In this section we provide the results of our experiments, focusing on our proposed model evaluation and how it performed versus the pre-trained models that we enforced on our study. Evaluation metrics that were used in our results were accuracy, F1-score, precision, recall and the confusion-matrix of our classification.

In our computational study we used both our multiple processed versions of our original dataset and gained various results and insights on our algorithms. To give a brief example, our algorithms were trained on the original dataset after some image resizing, on the grayscale transformation of our dataset as well as on an enriched training dataset with rotated versions of the images.

Our best result (see fig. 16) was a validation accuracy of 94.37% and was obtained when using our proposed CNN model architecture on a training dataset four times larger than the original, given that we added rotated angles of 90,180 and 270 degrees on the original images. We also downsized the original images to a size of 256x256.

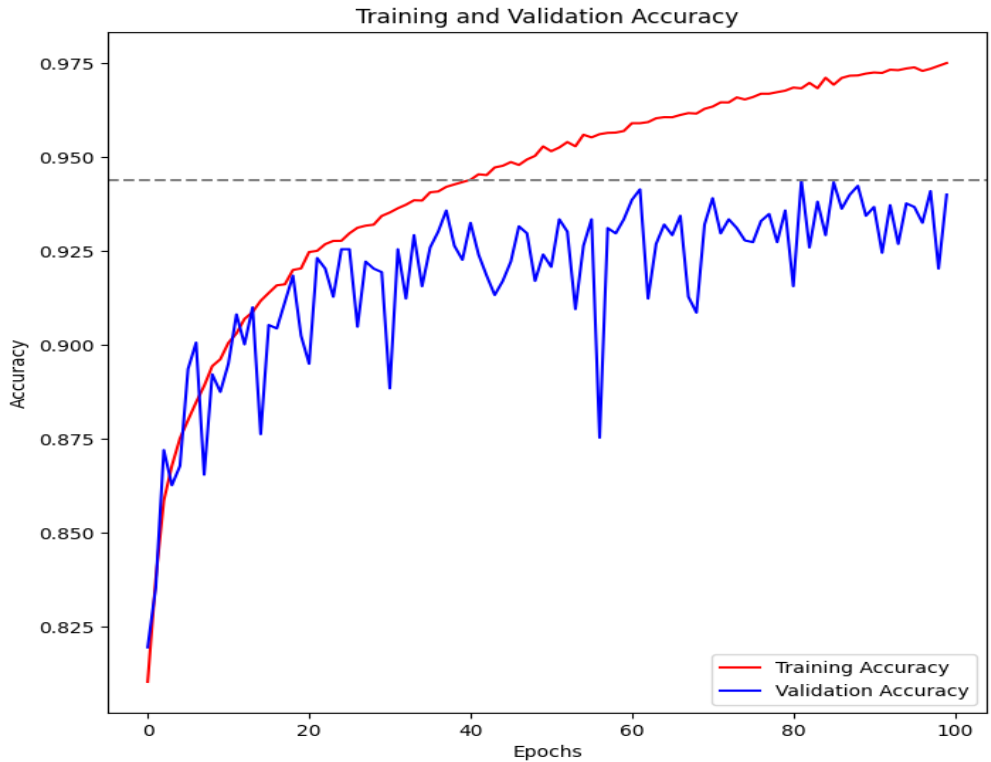

*Figure 16 Training Vs Validation Accuracy of Proposed Model*

As shown in figure 16, our algorithm begun to converge after a number of 70 epochs. We note that to help the algorithm converge and in order to improve accuracy and have a lower accuracy variance we changed the learning rate from 0.001 to 0.0001.

Evaluating our model there were 2005 correct predictions in 2133 testing images for an accuracy of 94.05 %. To visualize our results of our model better we present in fig. 17 the confusion matrix for our binary classification problem.

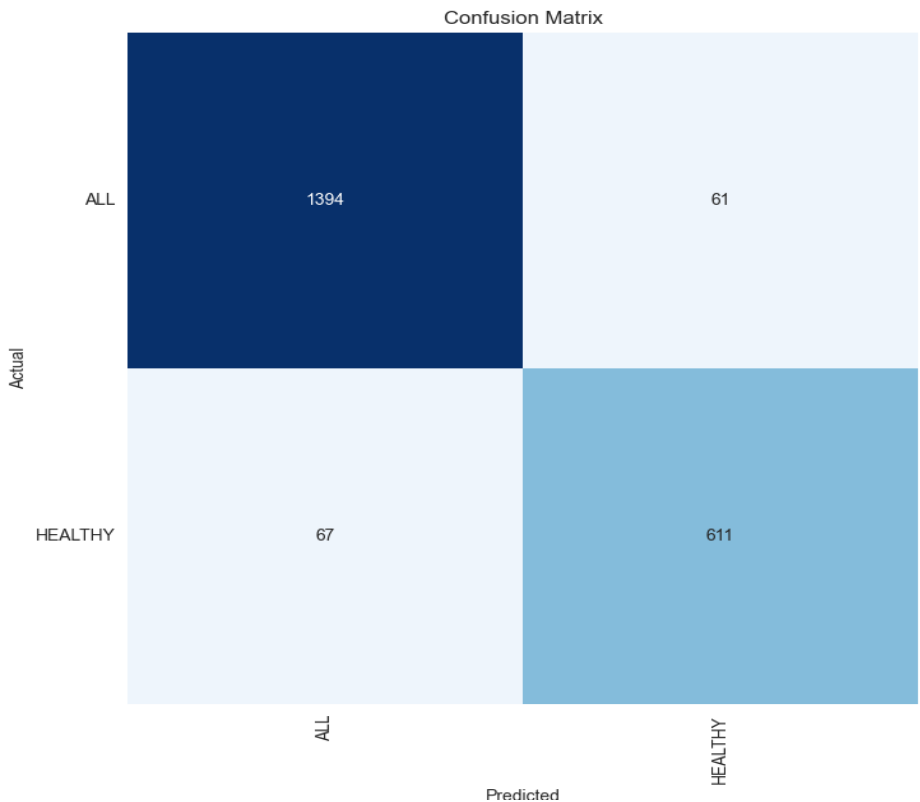

*Figure 17 Confusion Matrix of the proposed on the binary ALL/Healthy image cell classification*

What we can derive from the confusion matrix of fig. 17 is the following

- True Positive (TP): 1394/1445 ALL (cancer) image cells were correctly predicted as ALL.
- True Negative (TN): 611/678 HEALTHY image cells were correctly predicted as HEALTHY.
- False Positive (FP): 67/678 image cells predicted as ALL (cancer) that were actually HEALTHY.
- False Negative (FN): 61/1445 ALL (cancer) image cells were predicted as HEALTHY that were actually ALL (cancer).

Precision, recall, and F1 score are all important metrics for evaluating the performance of a machine learning model, particularly in classification tasks, by taking into account both the model's ability to make correct predictions (precision) and its ability to identify all actual positive observations (recall).

In order to provide a comprehensive assessment of our model performance we provide in Table 3 the evaluation metrics towards our two classes, as well as the macro evaluation metrics.

*Table 3 Evaluation Metrics of our Proposed Model on the image cell binary Classification (ALL/HEALTHY)*

|  | Precision | Recall | F1-score |
|---|---|---|---|
| ALL | 0.95 | 0.96 | 0.96 |
| HEALTHY | 0.91 | 0.90 | 0.91 |
|  |  |  |  |
| Macro average | 0.93 | 0.93 | 0.93 |
| Weighted average | 0.94 | 0.94 | 094 |

As shown in the upper table, the metrics on each class can be more informative than the macro metrics, as the macro metrics do not take the class imbalance into consideration. Since there is a small imbalance in our dataset, using the weighted average F1 score can help to better evaluate the performance of a classifier, as it gives more weight to the minority class, and it better reflects the importance of correctly classifying the minority class.

From the results we obtained we feel confident to assume that our classifier has a better prediction capability towards the ALL (cancer) class.

In image classification for cancer diagnosis, correctly identifying cancerous cells is generally considered more important than correctly identifying healthy cells. This is because misdiagnosis of cancer can have serious consequences, such as delayed treatment or unnecessary biopsies, while a healthy cell misdiagnosis may not have such severe consequences.

Before reaching the best result scenario that we described above, we tested our proposed model architecture on various image sizes and transformations on our original dataset which appear in Table 4.

*Table 4 Evaluation Results of the proposed model on different versions of the original dataset*

| Image Size | Color Format | Execution Time (min) | Training Dataset | Training size (images No) | Accuracy |
|---|---|---|---|---|---|
| 256X256 | RGB | 460 | 4x Rotated | 34116 | 94.37% |
| 128X128 | RGB | 67 | 4x Rotated | 34116 | 93.9% |
| 256X256 | RGB | 52 | Original | 8529 | 90.1% |
| 128X128 | RGB | 16 | Original | 8529 | 89.7% |
| 128X128 | Grayscale | 13 | Original | 8529 | 85.4% |

As shown in Table 4, we begun evaluation of our proposed model on the grayscale transformation of our dataset. As we moved further the model's accuracy increased when we introduced RGB channel images instead of Grayscale, feeding more information to our model. Eventually, adding rotated images to our training set, making our architecture more robust and unbiased towards the orientation of the images led to a significant accuracy boost to the maximum of 94.37%.

Finally, we compared the results of our proposed model with the performance of the pre-trained models Resnet-50 and MobileNetV2 (see fig. 18).

To clarify our work in a better way, it is important to underline that we did not train the pre-trained model's layers. So, what we did was what is called "freeze" the layers, so weights of those layers do not get updated. The pre-trained model has already learned useful features from a large dataset (ImageNet), so our aim was to check how these models would perform on our dataset. The only layers we added were a flatten layer followed by a dropout layer of a 20% dropout, a fully connected layer of 128 kernels and finally a Softmax output layer corresponding to our binary classification problem.

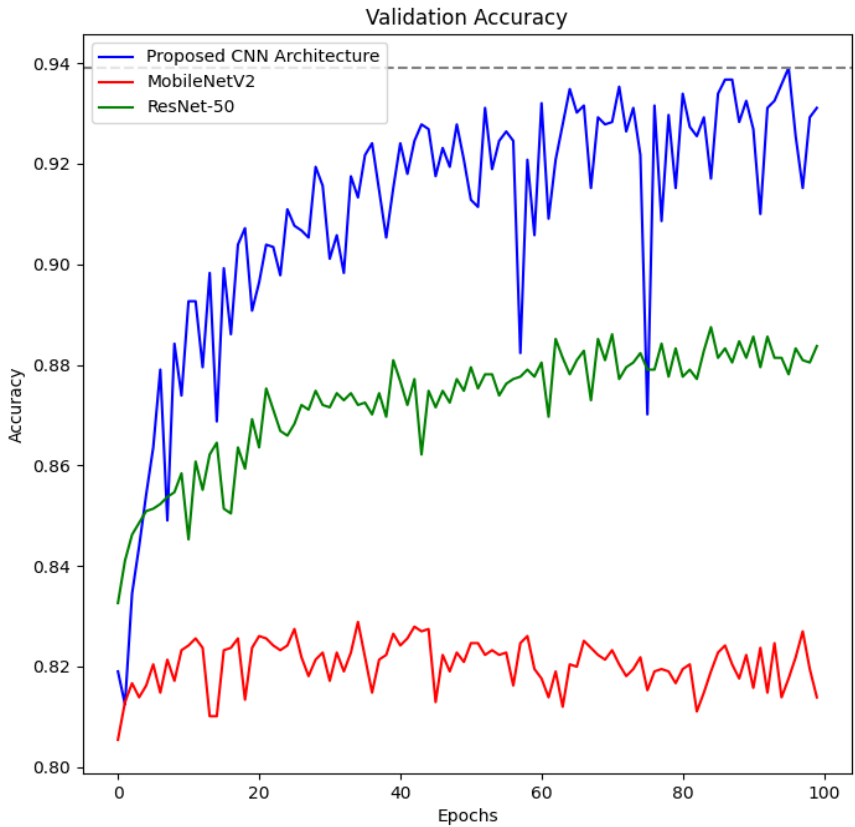

*Figure 18 Accuracy Evaluation of the proposed model versus pre-trained models (Resnet-50, MobilNetV2)*

To sum up the section of our results, we evaluate our three models towards accuracy (fig. 18), comparing their performance on the same training and validation sets.

All three models were tested on RGB images of the size 128X128 with a training set enhanced by their rotated copies on the angles of 90,180 and 270 degrees. From the accuracy graph it is straight-forward to say that our proposed model outperformed the pre-trained models for the task at hand of the binary image cell classification problem.

*Table 5 Comparing Models Results on the Same Data*

| Model | Image Size | Training size (images No) | Color Format | Execution Time (min) | Accuracy |
|---|---|---|---|---|---|
| Proposed CNN Model | 128X128 | 34116 | RGB | 67 | 93.9% |
| ResNet-50 | 128X128 | 34116 | RGB | 175 | 88.7% |
| MobileNetV2 | 128X128 | 34116 | RGB | 73 | 82.8% |

As we can see in Table 5, our proposed model reached an accuracy of 93.9% during the comparison experimentation of the models, when the other pre-trained models performed worse towards our image cell classification test. However, it is very important to underline that ResNet50 and MobileNetV2 were trained on the ImageNet dataset, with 14 million of various photos that consist around 1000 different classes. For the models to reach an accuracy of 89% (Resnet-50) without any real training on our own dataset goes to show why these models are considered well-tuned and good tools for image classification in general. Further customizing these pre-trained models and adding custom layers at their output would possibly produce an even better performance.

# Conclusions & Future Directions

A combination of heterogeneous morphologies, small dataset size, lack of annotated data, technical challenges make ALL image classification a difficult task. However, advances in deep learning and computer vision are helping to address these challenges and improve the accuracy of ALL image classification models. In this work, convolutional neural network (CNN) architectures were developed to classify cell images taken from microscopes in order to detect Acute lymphoblastic Leukemia (ALL). The goal of this study was to evaluate the performance of the CNN in classifying cancer images accurately.

Data acquisition was another significant part of the work presented in this thesis. As described in previous sections, getting access to a large dataset of image cells that can be useful and have a reasonable number of files to be valuable in a Deep Learning study, also considering the sensitive status that characterizes medical data, can prove really challenging.

Data pre-processing constituted a major factor towards our purpose of classifying cancer images with accuracy. Pre-processing techniques such as image resizing and grayscale transformation were used to give initial insights to our model structure, simplify our problem and make it more feasible given our system's hardware limitations. Furthermore, data augmentation techniques like image rotation were implemented that contributed in a significant accuracy boost of our result model.

The results of this study showed that the CNN model developed was able to accurately classify cancer images with an overall accuracy of 94.37%. This performance was found to be comparable to other well-performing models in the field.

There were several limitations to this study, including the limited diversity of the cancer images, so it is possible that the model may not generalize well to other types of cancer images. System limitations are another major bottleneck: training DL models requires a vast amount of computational power. Experimenting on a relatively old system, we were obligated to spend a long amount of time in each training iteration, especially when the model architectures we were enforcing, consisted of a large number of layers and the depth of them was quite demanding.

In conclusion, this work demonstrated the feasibility of using CNNs for ALL classification. It was shown through multiple experimental tuning on the CNN features, that the CNN model developed was able to accurately classify ALL image cells and had potential for further improvement. This study re-affirms the importance of deep learning in medical imaging and has potential implications for future studies in this area.

Regarding future work, our main focus would be on raising the performance of the CNN model by using larger and more diverse datasets, trying different CNN architectures, further experimenting on the model's hyperparameters and incorporating additional features or pre-processing techniques to enhance image classification accuracy.